\documentclass{ifacconf}

\usepackage[round]{natbib} 
\usepackage[T1]{fontenc}
\usepackage[latin9]{inputenc}
\usepackage{graphicx}

\makeatletter
\newcommand{\lyxdot}{.}

\begin{document}
\begin{frontmatter}

\title{Existence of Quasi-stationary states at the Long Range threshold}


\author[First]{Alessio Turchi}
\author[Second]{Duccio Fanelli}
\author[Third]{Xavier Leoncini}

\address[First]{Dipartimento di Energetica \textquotedbl{}Sergio Stecco\textquotedbl{},
Universita' di Firenze, via s. Marta 3, 50139 Firenze, Italia  and
Centre de Physique Théorique, Aix-Marseille Université, CNRS, Luminy,
Case 907, F-13288 Marseille cedex 9, France.}
\address[Second]{Dipartimento di Energetica \textquotedbl{}Sergio Stecco\textquotedbl{},
Universita' di Firenze, via s. Marta 3, 50139 Firenze, Italia and
Centro interdipartimentale per lo Studio delle Dinamiche Complesse
(CSDC) and INFN}
\address[Third]{Centre de Physique Théorique, Aix-Marseille Université, CNRS, Luminy,
Case 907, F-13288 Marseille cedex 9, France (e-mail: leoncini@cpt.univ-mrs.fr)}

\begin{abstract}                
In this paper the lifetime of quasi-stationary states (QSS) in the
$\alpha-$HMF model are investigated at the long range threshold ($\alpha=1$).
It is found that QSS exist and have a diverging lifetime $\tau(N)$
with system size which scales as $\mbox{\ensuremath{\tau}(N)\ensuremath{\sim}}\log N$,
which contrast to the exhibited power law for $\alpha<1$ and the observed
 finite lifetime for $\alpha>1$. Another feature of the long range
nature of the system beyond the threshold ( $\alpha>1$)  namely a phase transition
is  displayed for $\alpha=1.5$.  The definition of  a long range system is as well  discussed.
\end{abstract}

\begin{keyword}
Long range systems, Fractional dynamics, Hamiltonian chaos.
\end{keyword}

\end{frontmatter}


\section{Introduction}

Studying the dynamics of Hamiltonian systems with a large number of
degrees of freedom and its connection to equilibrium statistical mechanics
has been a long standing problem. The relaxation to statistical equilibrium
has been under scrutiny ever since the pioneering work of Fermi and
the FPU problem\cite{FPU55}. Moreover, since the advent of powerful
computers and for specific systems within a class of initial conditions,
integrating numerically Hamiltonian dynamics has proven to be competitive
in regards to Monte-Carlo schemes for the study of statistical properties
(see for instance \cite{Leoncini98,Torcini99} and references therein). The assumption made
is that since the system admits only a few conserved quantities for
generic initial conditions, once the dimensions of phase space are
large enough, microscopic Hamiltonian chaos should be at play and
be sufficiently strong to provide the foundation for the statistical
approach within the micro-canonical ensemble. However recent studies
have shown that there is an increase of regularity with system size
in the microscopic dynamics when considering systems with long range
interactions \cite{Chavanis-RuffoCCT07,Bachelard08,Leoncini09b,Vandenberg10}.
Indeed, the statistical and dynamical properties of these systems
are still under debate. For instance extensivity is not always provided
and discrepancies between canonical and micro-canonical ensembles can
be found such as negative specific heats for the latter \cite{Dauxois_book2002,Campa09}.
Moreover, phase transitions for systems embedded in one dimension can
be found. 

In particular, long range systems often display a slow relaxation
to equilibrium. Starting from an initial condition they are in fact
trapped in long-lasting out of equilibrium regimes, termed in the
literature Quasi Stationary States (QSS) which have distinct macroscopic
characteristics, when compared to the equilibrium configuration.

A now paradigmatic model of long range interactions Hamiltonian systems
is the Hamiltonian Mean Field (HMF) model \cite{Antoni95}, which
corresponds to a mean field $XY$-model with a kinetic energy term
(rotators). In the limit of infinite system size the HMF model can
be described using a Vlasov equation \cite{Chavanis2006,Campa09}.
More recently, stationary states have been constructed using invariant
measures of systems composed of uncoupled pendula \cite{Leoncini09b},
more specifically it was emphasized that the microscopic dynamics
in the magnetized stationary state is regular and explicitly known.
This observation lead to explain the abundance of regular orbits as
revealed in \cite{Bachelard08}. These results first obtained for
the HMF have been extended for the case when the coupling constant
depends on the distance between sites, namely for $\alpha-$HMF model
in its long range version ($\alpha<1$)\cite{Vandenberg10}. This
model was introduced for instance in \cite{Anteneodo98} and and displays
identical equilibrium features as the HMF\cite{Tamarit00,Campa09}.
In fact it was shown that all stationary states of the HMF model are
as well stationary states of the $\alpha-$HMF model, that microscopic
dynamics is as well regular, at the price of microscopic spatial complexity,
which is locally scale invariant\cite{Vandenberg10}. Before going
on we write the governing Hamiltonian of the model:

\begin{equation}
H=\sum_{i=1}^{N}\left[\frac{p_{i}^{2}}{2}+\frac{1}{2\tilde{N}}\sum_{j\ne i}^{N}\frac{1-\cos\left(q_{i}-q_{j}\right)}{\Vert i-j\Vert^{\alpha}}\right]\:,\label{eq:Hamiltonian_HMF}\end{equation}
where $q_{i}$ stands for some spin angle located on the lattice site
$i,$ and $p_{i}$ is its canonically conjugate momentum. The distance
$\Vert i-j\Vert$ is actually the shortest distance on the circle
of perimeter $N-1$, so that the systems can be isolated and still
translational invariant along the lattice. The mean field model is
recovered for $\alpha=0$, and for $N$ even, we write \begin{equation}
\tilde{N}=\left(\frac{2}{N}\right)^{\alpha}+2\sum_{i=1}^{N/2-1}\frac{1}{i{}^{\alpha}}\:,\label{eq:Tilde_N}\end{equation}
to insure extensivity. The equations of motions of element $i$ are
derived from the Hamiltonian (\ref{eq:Hamiltonian_HMF}):\begin{eqnarray}
\dot{p_{i}} & = & -\sin(q_{i})C_{i}+\cos(q_{i})S_{i}=M_{i}\sin(q_{i}-\varphi_{i})\:,\label{eq:p_dot_bis}\\
\dot{q_{i}} & = & p_{i}\:,\end{eqnarray}
where \begin{eqnarray}
C_{i} & = & \frac{1}{\tilde{N}}\sum_{j\ne i}\frac{\cos q_{j}}{\Vert i-j\Vert^{\alpha}}\label{eq:C_simple_N_tilde}\\
S_{i} & = & \frac{1}{\tilde{N}}\sum_{j\ne i}\frac{\sin q_{j}}{\Vert i-j\Vert^{\alpha}}\:.\label{eq:S_simple_N_tilde}\end{eqnarray}
$C_{i}$ and $S_{i}$ identify the two components of a magnetization
per site, with modulus $M_{i}=\sqrt{C_{i}^{2}+S_{i}^{2}}$ , and phase
$\varphi_{i}=\arctan(S_{i}/C_{i})$. For large $N$ , and assuming
$0<\alpha<1$, we have

\begin{equation}
\tilde{N}\approx\frac{2}{1-\alpha}(N/2)^{1-\alpha}\:.\label{eq:N_tilde_large_N}\end{equation}
We then can use (\ref{eq:N_tilde_large_N}) in Eq.(\ref{eq:C_simple_N_tilde})
and, make the $N\rightarrow\infty$ limit while introducing the continuous
variables $x=i/N$ and $y=j/N$ to arrive at 

\begin{equation}
C(x)=\frac{1-\alpha}{2^{\alpha}}\int_{-1/2}^{1/2}\frac{\cos\left(q(y)\right)}{\Vert x-y\Vert^{\alpha}}dy\:,\label{eq:C_of_x}\end{equation}
where $\Vert x-y\Vert$ represents the minimal distance on a circle
of perimeter one. We can recognize the fractional integral $I^{1-\alpha}$and
consequently write \begin{equation}
C(x)=\frac{1-\alpha}{2^{\alpha}}\Gamma(1-\alpha)I^{1-\alpha}\left(\cos q(x)\right)\:.\label{eq:Fractional_C}\end{equation}
 In this large size limit, the $\alpha-$HMF dynamics implies studying
the evolution of the scalar fields $q(x,t)$ and $p(x,t)$ which are
ruled by the fractional (non-local) partial differential equations\begin{eqnarray*}
\frac{\partial q}{\partial t} & = & p(x,t)\\
\frac{\partial p}{\partial t} & = & \frac{\mu}{2^{\alpha}}\Gamma(\mu)\left(-\sin(q)I^{\mu}\left(\cos q\right)+\cos(q)I^{\mu}\left(\sin q\right)\right)\:.\end{eqnarray*}
where $\mu=1-\alpha$. It has then been shown in \cite{Vandenberg10}
that stationary states are solutions of \begin{equation}
{\cal D}^{\alpha}\cos q=\frac{d^{\alpha}\cos q}{dx^{\alpha}}=0\:.\label{eq:fractional_Steady_state}\end{equation}
where the operator $D^{\alpha}$ stands for the fractional derivative,
and that actually this property was shared with non stationary QSS's.
All these results were obtained for the model in its long range version,
meaning when $\alpha$ is smaller than one. 

\section{What happens for $2>\alpha\ge1$}

\begin{figure}
\begin{centering}
\includegraphics[width=7cm]{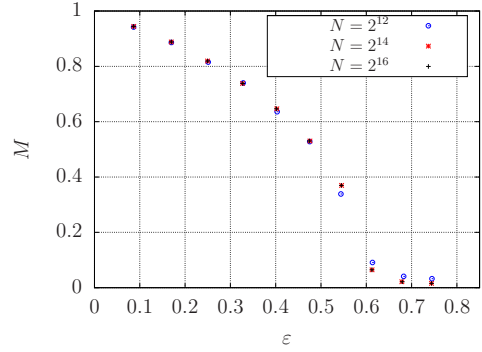} 
\par\end{centering}

\caption{%
Magnetization vs energy, for $\alpha=1.5$. A phase transition is
displayed with $\epsilon_{c}\approx0.6$. The transition point seem
to be different from the long-range one which is $\epsilon_{c}=0.75$,
but the qualitative behavior of curve is the same.\label{fig:Magnetization-vs-energy,} 
}

\end{figure}

We recall that systems
are considered long range when the two body interacting potential
$V(r)$ decays at the most as $1/r^{\alpha}$ with $\alpha<d$, where
$d$ stands for the dimension of the embedding space. Having only one degree
of freedom $d=1$ for the $\alpha-$HMF model.  In these regards,
considering situations where  $1<\alpha<2$ is actually studying short range models,
In fact  $\tilde{N}$ is  finite so there is no need of system size renormalization of the couling constant
for $\alpha>1$.
However something is peculiar about this lattice model. Indeed when
considering the dynamics of a long range system, we would expect the
force to be ruled by a $1/r^{\beta}$ decay with $\beta=\alpha+1$,
which is not the case for the lattice model for which the decay exponent
is unchanged ($\beta=\alpha$). Moreover given the particular importance
of the microscopic dynamics and possible ergodicity breaking, one
could naturally raise the question if the long range nature of a system
is not ruled by the dynamics, which would then imply a system to be
long ranged if $\beta<d+1$, which for the $\alpha-$HMF model would
imply $\alpha<2$. Most of the previous analysis of the model has
been performed for $\alpha<1$, and can not simply be extended to
$1<\alpha<2$. A first numerical analysis is therefore necessary.

\subsection{Phase transitions}

One peculiarity of one dimensional systems, is that there should not
bare any phase transition if the interaction is short ranged. A first
numerical study of the magnetization versus density of energy is performed
for $\alpha=1.5$ in Fig.~\ref{fig:Magnetization-vs-energy,}. The
numerical integration of the microscopic dynamics is performed using
a simplectic scheme, (optimal fifth order see\cite{McLachlan92}),
a typical time step used is $\delta t=0.05$, and the initial conditions
are Gaussian distributed. The fast Fourier transforms are done with
the fftw libraries. As can be seen in Fig.~\ref{fig:Magnetization-vs-energy,}
a phase transition is displayed at a transition point $\epsilon_{c}\approx0.6$
which is then different from the universal value obtained for $\alpha<1$
which is $\epsilon_{c}=0.75$. Preliminary results show actually that
the critical point depends on the value of $\alpha$ and results seem
to show that it approaches $\epsilon=0$ for $\alpha=2$. 

\begin{figure}
\begin{centering}
\includegraphics[width=8cm]{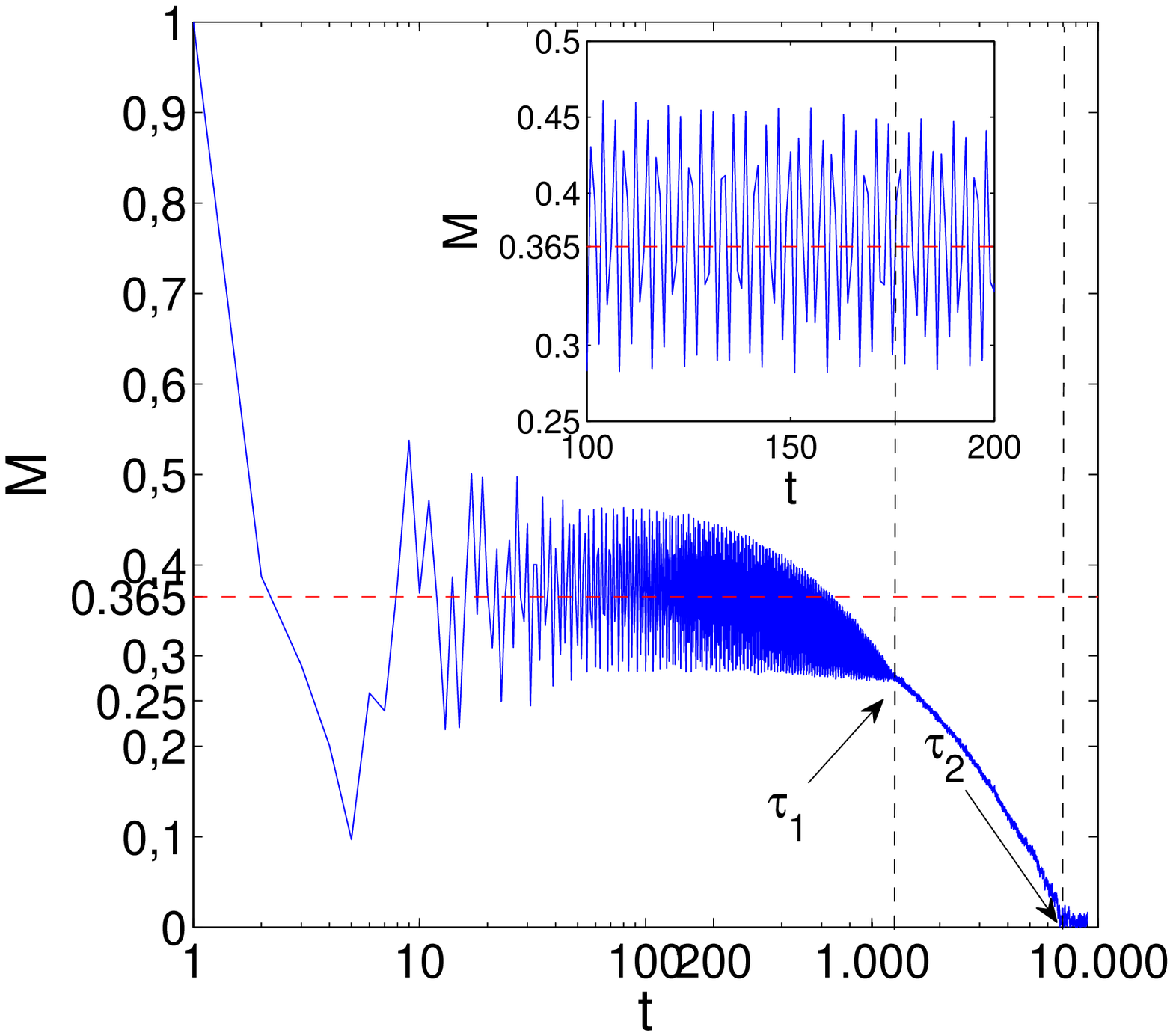}\\
 \includegraphics[width=8cm]{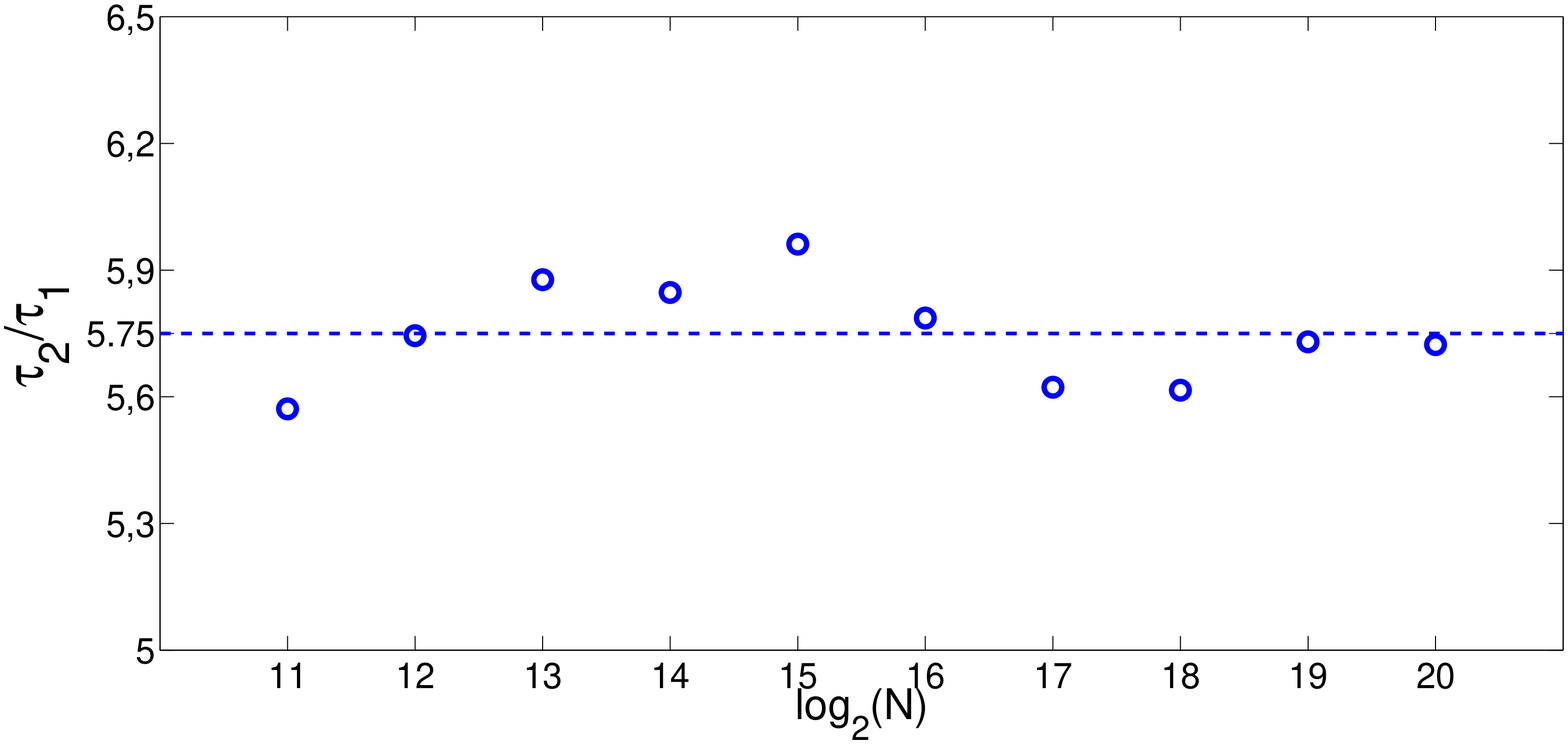} 
\par\end{centering}

\caption{%
Top: Magnetization curve vs time for $\alpha=1.0$, $N=2^{1}8$ and
$\epsilon=1.2$. During the QSS the magnetization is oscillating for
a time $\tau_{1}$, then it relaxes down to it's equilibrium value
in a time $\tau_{2}$.\protect \\
 Bottom: $\frac{\tau_{2}}{\tau_{1}}$ for different values of $N$,
$\alpha=1$ and $\epsilon=0.6$. The two times are approximately proportional,
so the knowledge of $\tau_{1}$ gives a good approximation for $\tau_{2}$.
This proportionality is respected for all $\alpha$ values in this
paper, even if the proportionality constant may vary. \label{fig:Top:-Magnetization-curve}
}

\end{figure}
This existence of a phase transition beyond the {}``classical''
long range threshold in one dimension had already been noticed for
the Ising model by Dyson in the sixties, but this feature seems
to favor the dynamical definition of what a long range system ought
be. However an important feature to assert this new definition would
be to find as well quasi-stationary states in this region of $\alpha$'s.

\subsection{QSS lifetime}

In long range interacting systems, generally, the limit $N\rightarrow\infty$
and $t\rightarrow\infty$ doesn't commute, so the thermodynamic limit
is not unequivocally defined and one may end up in different equilibrium
states depending on the order of the previous limits. Physically it's
more feasible to compute the continuous limit before the time limit,
so if the lifetime $\tau$ of the QSS diverges with $N$ it becomes
the effective real equilibrium of the system, which in general is
not obeying Boltzmann's statistics \cite{Chavanis-RuffoCCT07}. We
studied how the QSS lifetime scales with the exponent $\alpha$ in
the decay parameter of the potential. 

First we studied the behavior of the lifetime $\tau$ around the
crucial value $\alpha=1$ to better understand the transition between
a long range system and a supposedly short range one, but as mentioned
in the case of $\alpha$-HMF there are convincing arguments that the
requirement for the emergence of long range behavior could be relaxed,
and we can expect some long range feature to survive above $\alpha=1$.%
\begin{figure}
\begin{centering}
\includegraphics[width=7cm]{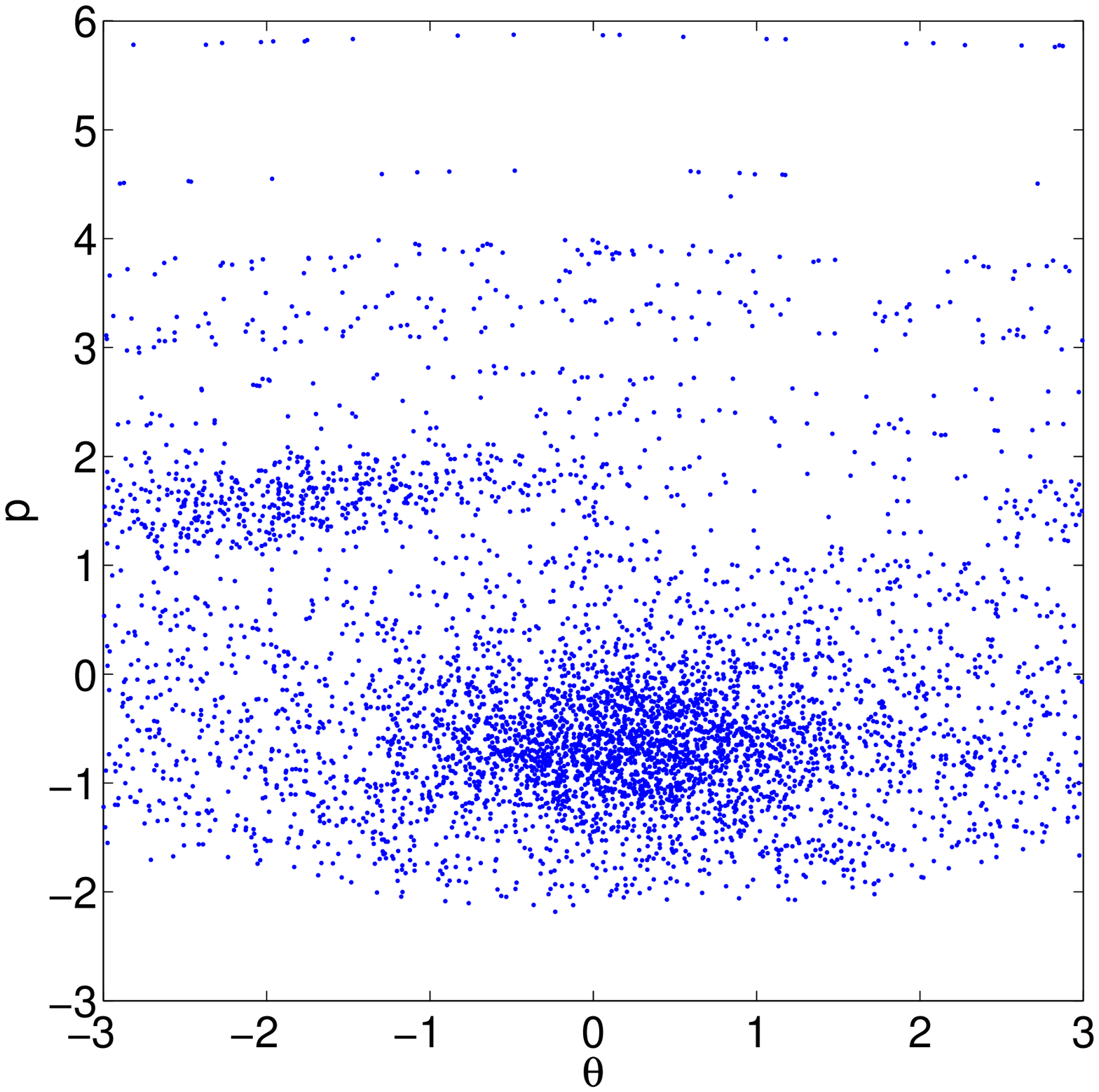}\\
 \includegraphics[width=7cm]{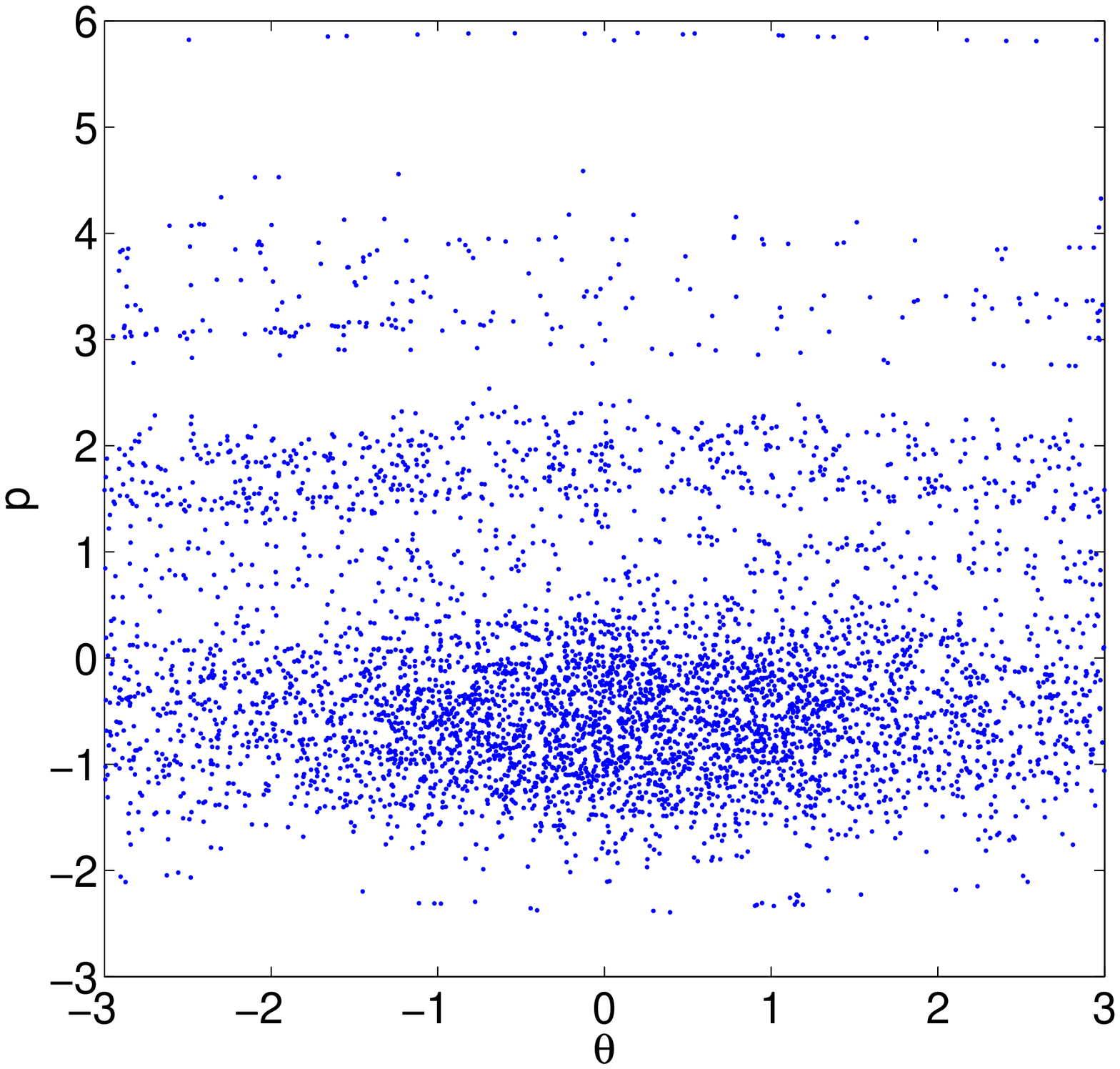} 
\par\end{centering}

\caption{%
Poincaré sections for $\alpha=1$, $\epsilon=1.2$ and $N=2^{20}$.
Top refers to the QSS, while bottom represent the relaxation state
between $\tau_{1}$ and $\tau_{2}$. It can be easily seen that the
second small isle disappear during the relaxation and the phase space
becomes symmetric in $q$, thus ending the oscillations of the magnetization.
\label{fig:Poincar=0000E9-sections-for}
}

\end{figure}

We considered the initial condition already used in \cite{Vandenberg10}
which was giving rise to a QSS, namely a long lived magnetized state
above the critical energy. The initial condition used is all $q_{i}=0$,
wile the $p_{i}$'s are Gaussian. To characterize the QSS lifetime
we monitor the behavior of the global macroscopic parameter magnetization
which can be $M=|\sum_{j}e^{iq_{j}}|$. In Fig.~\ref{fig:Top:-Magnetization-curve}
we show the behavior for $\alpha=1$, which is qualitatively representative
for the all studied values of $\alpha$ in this paper. %
\begin{figure}
\begin{centering}
\includegraphics[width=7cm]{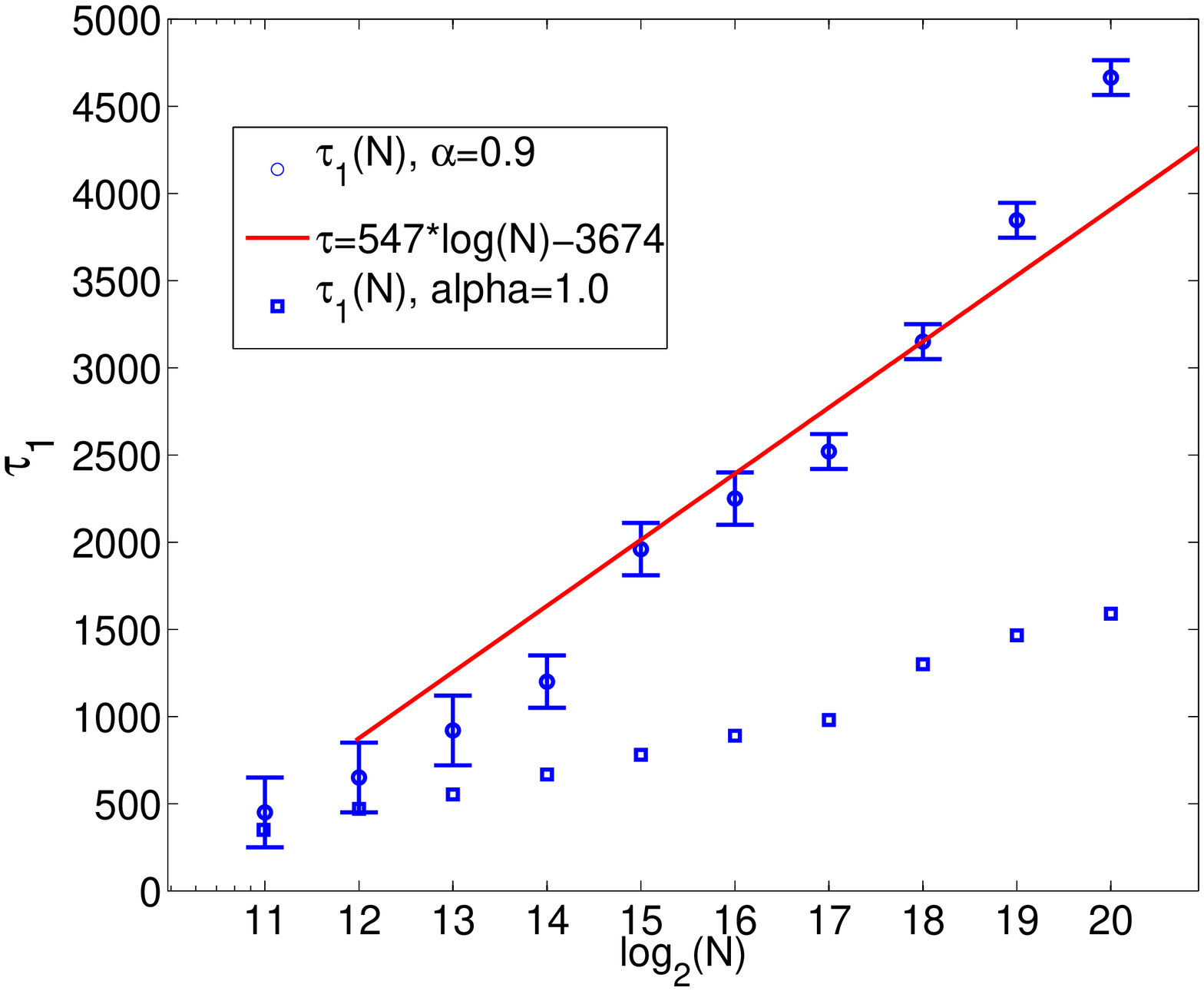}
\includegraphics[width=7cm]{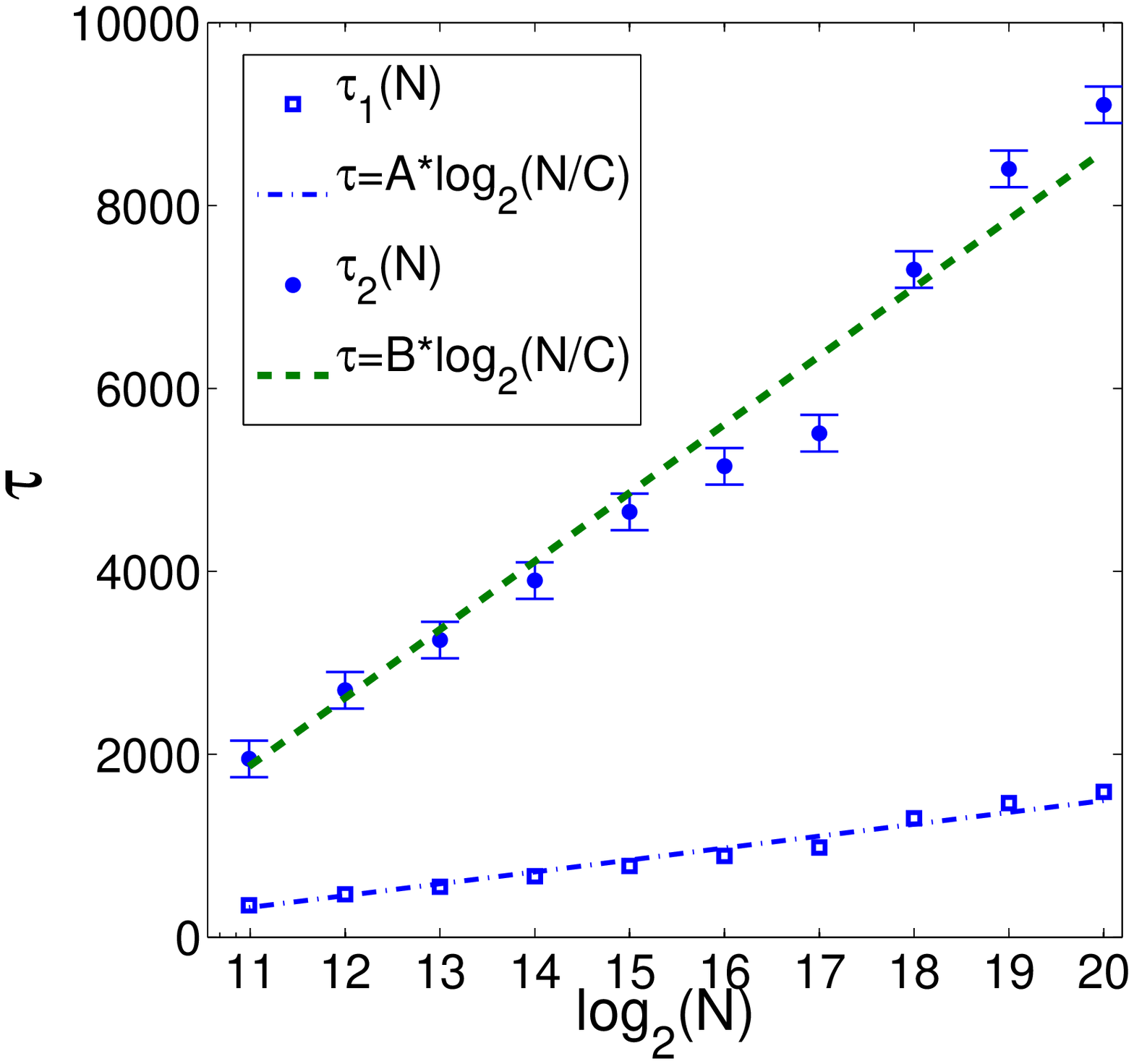}\\
 \includegraphics[width=7cm]{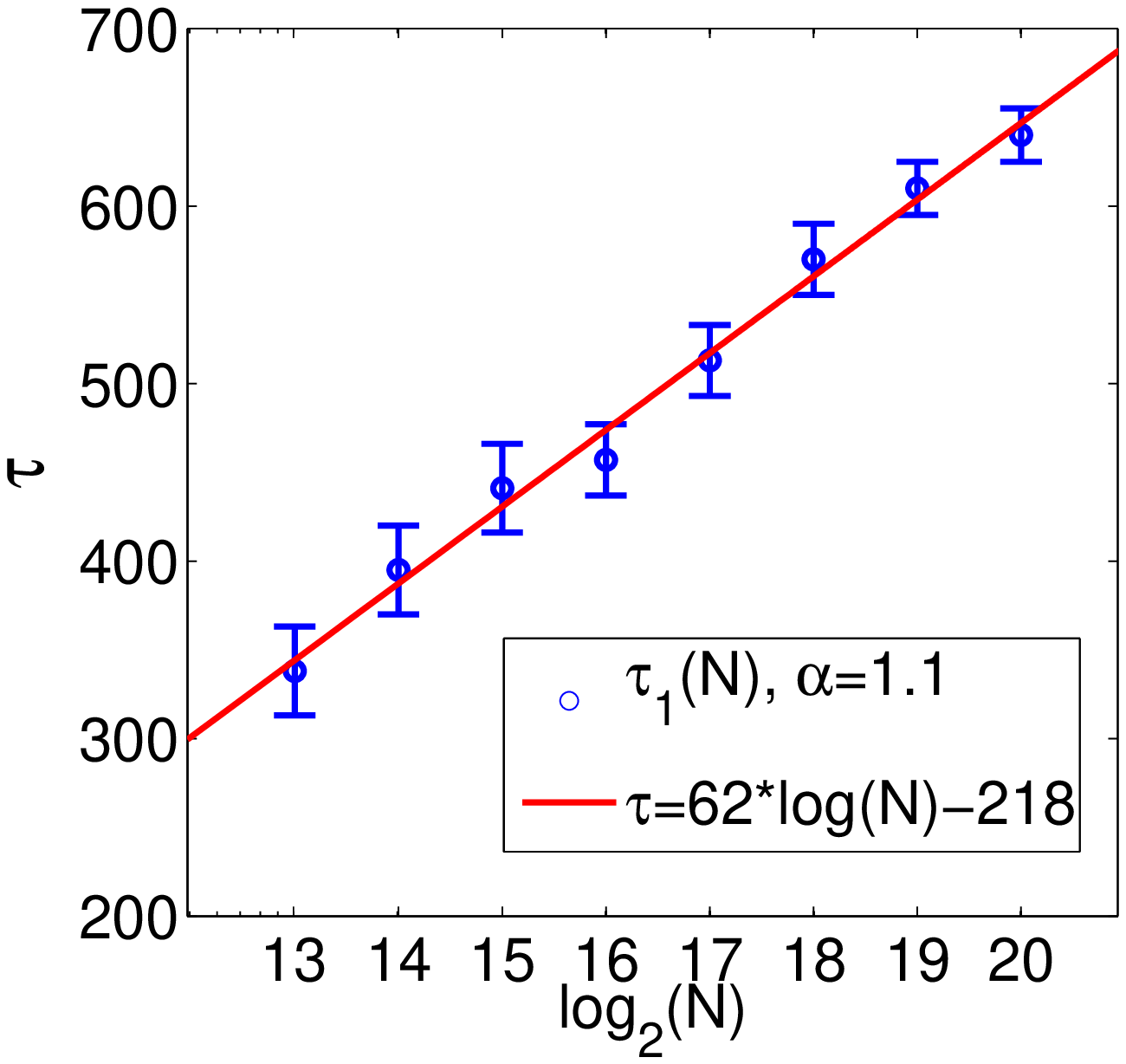}
\par\end{centering}

\caption{
Scaling of $\tau_{1}$ with $N$ for different $\alpha$ values and
$\epsilon=1.2$. Top shows the scaling for $\alpha=0.9$; middle refers
to $\alpha=1.0$ and shows both $\tau_{1}$ and $\tau_{2}$; while
bottom refers to $\alpha=1.1$ . All curves seems to grow at least
logarithmically. \label{fig:Scaling-of-}
}

\end{figure}
 Here a first transition at $t=\tau_{1}$ can be identified until
which the system oscillates around an almost constant magnetization
value, and beyond which the system starts to relax towards the equilibrium
$M=0$ value. A second transition at $t=\tau_{2}$ is as well identified,
it corresponds to the time at which the system finally reaches it's
equilibrium state. As can be seen in Fig.~\ref{fig:Top:-Magnetization-curve},
we find that these two values are almost linearly proportional for
each $\alpha$-value that we took into consideration, so we will refer
to the first lifetime $\tau_{1}$ as the lifetime of the QSS, since
is an order of magnitude faster to compute and we are interested only
in the qualitative form of the scaling law for the lifetimes.%
\begin{figure}
\begin{centering}
\includegraphics[width=7cm]{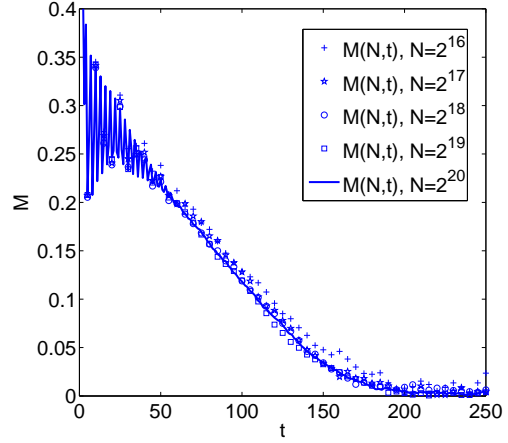} 
\par\end{centering}

\caption{
Magnetization curve vs time for $\alpha=1.5$, $\epsilon=1.0$ and
increasing $N$ values. The initial oscillation lifetime $\tau_{1}$
is independent of $N$ and the system relaxes to equilibrium in a
time which should be constant in the continuous limit.\label{fig:Magnetization-curve-vs}\protect \\
}

\end{figure}

The difference of the two dynamical regimes defined by the above thresholds
are better understood when looking at the Poincaré section captured
in each of this regimes displayed in Fig.~\ref{fig:Poincar=0000E9-sections-for}.
At first the system is forms two distinct islands in the phase space,
which start moving around and create the oscillations in the magnetization
that characterize the QSS, and then one of the isles disappear during
the relaxation period, while the phase space becomes symmetric in
$q$ thus ending the oscillations.

Now we analyze the lifetime of the QSS versus the size of the system
around the classical long range threshold $\alpha=1$, if the system
is long range it should diverge with $N$. The cases $\alpha=0.9$,
$\alpha=1$ and $\alpha=1.1$ are displayed in Fig.~\ref{fig:Scaling-of-}.
We can see that around $\alpha=1$ the scaling of $\tau$ with $N$
approaches a logarithmic curve, meaning that the QSS survive at least
until $\alpha=1$ and maybe beyond as it appears as well true for
$\alpha=1.1$. However when looking at the data for larger value of
$\alpha$, namely $\alpha=1.5$ the scaling of $\tau$ appears to
saturate as displayed in Fig.~\ref{fig:Magnetization-curve-vs},
where the magnetization curves obtained for $\alpha=1.5$ appear to
all be identical no matter the size of the system. At this value of
$\alpha$ there is still a short initial oscillation in $M$ typical
of the QSS, but now it's lifetime seems to be independent of $N$
and finite so the system will reach the equilibrium state in a large
enough time. Conversely we may expect that we may actually observe
the same feature for $\alpha=1.1$, but that we have not seen yet
the saturation in the lifetime as we were not able to simulate systems
that would be large enough. In other words, we have as now not enough
data to identify if there is a transition value between $1.1<\alpha<1.5$
where the system becomes suddenly short-range dropping the logarithmic
law of $\tau_{1}$, or it still saturates at some larger value of
$N>2^{20}$ even for $\alpha=1.1$. Preliminary results shows that
this saturation becomes quite fast for $\alpha\sim1.2,1.3$ where
even for low values of $N$ it may still be possible to observe a scaling which
is sub-logarithmic, but even if it did this would be only relevant  for
systems with an astronomical scale of constituants.

However even if the QSS lifetime seems to saturate at some point,
so that the system dynamics change into a short range one, the phase
transition from a magnetized to an homogeneous state, typical of the
long range regime, is still present (figure \ref{fig:Magnetization-vs-energy,}).
Hence is appears that the dynamical definition may be more relevant
for macroscopic features, such as the presence of a QSS (even with
a finite lifetime) or a phase transition, while the more classical
statistical definition of a long range system corresponds actually
to different dynamical behavior of the system and the existence of
QSS with diverging lifetimes.

\end{document}